\documentclass[twocolumn,aps,prd,amsmath,amssymb,showpacs]{revtex4}
\usepackage{epsfig,bm,dcolumn,color}
\usepackage{graphicx}
\usepackage{pstricks}

\newcommand {\beq} {\begin{equation}}
\newcommand {\eeq} {\end{equation}}

\newcommand {\up}  {\ensuremath{\uparrow}}
\newcommand {\dn}  {\ensuremath{\downarrow}}
\newcommand {\bqa} {\begin{eqnarray}}
\newcommand {\eqa} {\end{eqnarray}}

\newcommand {\kk} {\ensuremath{{\bf k}}}
\newcommand {\pp} {\ensuremath{{\bf p}}}
\newcommand {\QQ} {\ensuremath{{\bf Q}}}

\newcommand {\qq} {\ensuremath{{\bf q}}}

\begin{document}
\title{Universal Short-Distance Structure of the Single-Particle Spectral Function of Dilute Fermi Gases}
\author{William Schneider and Mohit Randeria}
\affiliation{Department of Physics, The Ohio State University, Columbus, Ohio 43210}
\begin{abstract}
We show that the universal $1/k^4$ tail in the momentum distribution of dilute Fermi gases implies that
the spectral function $A(\kk,\omega)$ must have weight below the chemical potential for 
large momentum $k \gg k_F$, with observable consequences in RF spectroscopy
experiments. We find that this incoherent spectral weight is centered about
$\omega \simeq - \epsilon(\kk)$ in a range of energies of order $v_F k$. This ``bending back'' in the dispersion,
while natural for superfluids, is quite surprising for normal gases.
This universal structure is present in the hard-sphere gas as well as the
Fermi liquid ground state of the highly imbalanced, attractive gas near unitarity.
We argue that, even in the BCS superfluid, this bending back at large $k$ is dominated by interaction
effects which do not reflect the pairing gap. 
\end{abstract}


\maketitle 

%%%%%%%%%%%%%%%%%%%%%%%%%%%%%%%%%%%%%%

The spectral function $A(\kk,\omega) = - {\rm Im}G(\kk,\omega + i0^+)/\pi$ of
the single-particle Green's function $G$ is 
of fundamental interest in many-body physics \cite{agd,kb}.
In addition to information about the spectrum and dynamics of single-particle
excitations, it is also directly related to thermodynamic functions of
a many-particle system.  
Very recently there has been experimental progress in measuring (the 
occupied part of) $A(\kk,\omega)$ in strongly interacting Fermi gases \cite{bloch_rmp,stringari_rmp}, 
using a momentum-resolved version \cite{jin2008} 
of radio frequency (RF) spectroscopy \cite{chin,ketterle2008,zwierlein2009}.
These measurements \cite{jin2008} of $A(\kk,\omega)$ for ultracold atomic gases are the analog of 
angle-resolved photoemission, which has given deep new insights into novel materials.

In this paper we uncover remarkable universal large-$k$ structure of $A(\kk,\omega)$
for dilute gases with observable effects in RF experiments. 
Our investigation was motivated by the elucidation of the 
universal ultraviolet structure of equal-time correlations
by Tan \cite{tan,others}. One of his central results is the
universal $k \gg k_F$ behavior of the momentum distribution 
$n_\sigma(\kk) \simeq C/k^4$, where $C$ is the ``contact'' \cite{tan,others}. Using the
$T=0$ sum rule $\int_{-\infty}^0 d\omega A(\kk,\omega) = n(\kk)$, this necessarily implies that
$A(\kk,\omega)$ has weight below the chemical potential ($\omega < 0$) for
$k \gg k_F$. This is ``incoherent'' spectral weight,
\emph{not} associated with the coherent Landau quasiparticle.

We ask the question: Where is this incoherent spectral weight located for $k \gg k_F$?
The surprising answer is that the incoherent part of the $\omega\ vs.\ k$ dispersion 
goes like $- \epsilon(\kk) = - k^2/2m$, ``bending back'' away from the chemical potential at large $k$.
While this is expected in BCS theory and its generalizations for
a paired superfluid, we argue that this unusual dispersion is a universal feature of 
\emph{all} dilute Fermi gases, even those with a \emph{normal} (non-superfluid) ground state.
We find that the spectral weight of $C/k^4$ in $A(\kk,\omega)$ is centered about 
$\omega \simeq - \epsilon(\kk)$ in a range of energies of order $v_F k$ for normal Fermi gases. 
Most of the spectral weight $(1 - C/k^4)$ is, of course, centered about
$\omega \approx + \epsilon(\kk)$, but these states are not occupied and 
do not contribute to $n(\kk)$. 

This bending back is clearly visible in the data of ref.~\cite{jin2008} for attractive
fermions near unitarity and near or above $T_c$. 
However, it is hard to separate the effects of the finite 
temperature pairing pseudogap \cite{varenna}
and normal state interaction effects. In particular,
a bending back of the dispersion above $T_c$ cannot by itself be used as evidence
for a pairing pseudogap in view of the normal state results described below.

We will first focus on two systems where the ground state
is a normal Fermi liquid: (a) the hard-sphere dilute Fermi gas, and (b) the highly imbalanced attractive Fermi gas. 
We then turn to the superfluid ground state,
where we will argue that, in the BCS limit, the unusual dispersion is dominated by interaction effects
rather than the effect of pairing. We conclude with
implications for RF spectroscopy experiments.

{\bf Dilute repulsive Fermi gas:} We begin with the 3 dimensional
hard sphere Fermi gas with dispersion $\epsilon(\kk) = k^2/2m$, mass $m$, density $n =k_F^3/ 3\pi^2$, and scattering length $a > 0$ with $na^3 \ll 1$. (We set $\hbar=k_B=1$.)
Its thermodynamic and Fermi-liquid properties were studied by
Galitskii and Lee, Yang and Huang; see Sec.~5 of \cite{agd}. The high-$k$ tail was also calculated \cite{belyakov,range}:   
$n(\kk) \simeq (k_F a)^2 \left(2/3\pi\right)^2\left(k_F/k\right)^4$. Here we compute $A(\kk,\omega)$.

In the low density limit $n a^3 \ll 1$, the most important physical process
is repeated scattering in the particle-particle channel. The corresponding
sum of ladder diagrams $\Gamma$ is given by $\Gamma^{-1}(Q) = 1/g - L(Q)$,
where $Q =(\QQ, iQ_\ell)$ with $iQ_\ell = i2\ell\pi T$ and
the bare interaction $g$ is related to $a$ via $1/g = m/(4\pi a) - \sum_\kk 1/[2\epsilon(\kk)]$.
Further $L(Q) = T \sum_k G^0(k+Q)G^0(-k)$ where $k=(\kk, i k_n)$ with $ik_n = i(2n+1)\pi T$ and
$G^0(k) = 1/[ik_n - \xi(\kk)]$ is the bare Green's function with the energy
$\xi(\kk) = \epsilon(\kk) - \mu$ measured with respect to the chemical potential $\mu$ \cite{mu-footnote}.
Note that one can obtain an analytically closed form expression \cite{nikolic} for $L(\QQ,\Omega + i0^+)$.  
For the hard sphere gas we can make a further simplification 
$\Gamma \approx g + g^2 L$.

%====================================================================================

\begin{figure}
\centerline{\epsfxsize=8.0truecm \epsfbox{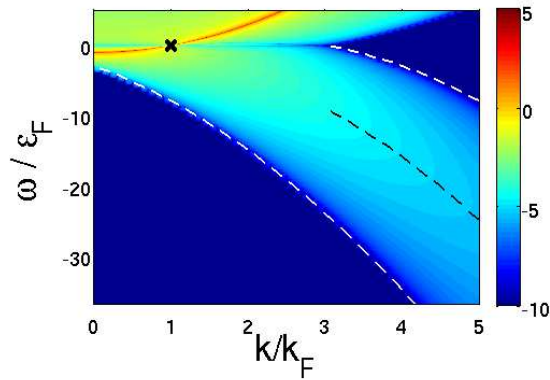}}
\caption{(color online) Logarithmic intensity plot of $A(\kk,\omega)\epsilon_F /(k_F a)^2$ for the repulsive Fermi gas
($k_F a = 0.1$; $na^3 = 3.4 \times 10^{-5}$). The most intense (red) line at $\omega \approx \xi_k$ is the quasiparticle.
We focus on the unusual dispersion centered about $\omega = - \epsilon(\kk)$
(black dashed line) in the range $\omega = - \epsilon(\kk) - 3\epsilon_F \pm 2v_F k$ (white dashed lines); see text.
}
\label{akw-galintskii}
\end{figure}

%====================================================================================

\begin{figure}[b]
\epsfxsize=8.0truecm \epsfbox{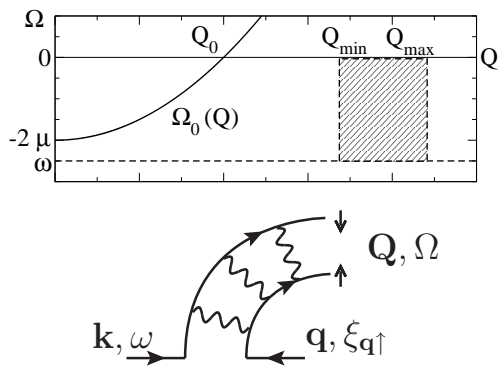}
\caption{(color online) Top: Kinematics of the processes that contribute to 
imaginary self-energy in eq.~(\ref{imsigma}). ${\rm Im}\Sigma$ is non-zero
when the shaded rectangle (allowed by kinematics and thermal factors) overlaps with the region 
$\Omega > \Omega_0(Q)$(in which ${\rm Im}\Gamma$ is non-zero). This leads to the condition
$Q_{\rm min} \le Q_0$. Bottom:
Diagram contributing to ${\rm Im}\Sigma$.}
\label{kinematics}
\end{figure}

The Matsubara self-energy $\Sigma(k) = T\sum_{q}\Gamma (k+q) G^0(q)$ yields
$\Sigma(\kk,ik_n \rightarrow \omega+i0^+) = {\rm Re}\Sigma + i{\rm Im}\Sigma$, where
\beq
{\rm Im}\Sigma(\kk,\omega) = \sum_\qq
{\rm Im}\Gamma(\QQ,\Omega)\left[\Theta\left(-\xi(\qq)\right) - \Theta(-\Omega)\right]
\label{imsigma}
\eeq
at $T=0$, with $\QQ = \kk + \qq$ and $\Omega = \omega + \xi(\qq)$.
${\rm Re}\Sigma$ is obtained numerically by a Kramers-Kronig transform \cite{KK} on ${\rm Im}\Sigma$.
The spectral function is then obtained using 
$A(\kk,\omega) = -{\rm Im}\left[\omega - \xi(\kk) - \Sigma(\kk,\omega)\right]^{-1}/\pi$
and plotted in Fig.~\ref{akw-galintskii} on a log scale.
We see the most intense feature, corresponding to the
Landau quasiparticle near $k_F$, tracks $\omega \approx + \xi(\kk)$, up to
many-body renormalizations \cite{checks}. 
However our main interest is in the much less intense, incoherent spectral feature 
that follows an $\omega = - \epsilon(\kk)$ dispersion and dominates $n(\kk)$ at large $k$.

To understand this ``bending back'', we write
$A \approx |{\rm Im}\Sigma(\kk,\omega)|/\left[\pi(\omega - \epsilon(\kk))^2\right]$.
We need to determine \emph{when} ${\rm Im}\Sigma(k,\omega)$ \emph{is non-zero}
for $k\gg k_F$ and $\omega<0$.
To understand our result qualitatively, consider the diagram in Fig.~\ref{kinematics}.
The dominant contribution comes from small values of both $|\QQ|$ and $\Omega$. (For large values
of these variables there is no spectral weight ${\rm Im}\Gamma$ for two-particle scattering.) Thus
$\qq \simeq - \kk$ and $\omega \simeq - \xi (\qq) \simeq - \epsilon(\kk)$ for $k \gg k_F$.
This shows that $A \neq 0$ for $\omega$ around \emph{negative} $\epsilon(\kk)$.

To make this more quantitative, we use eq.~(\ref{imsigma}).
From the structure of $L(\QQ,\Omega)$, it follows that ${\rm Im}\Gamma(\QQ,\Omega) \neq 0$
when $\Omega \ge \Omega_0(Q) \equiv \min_\pp \{\xi(\pp+\QQ/2) + \xi(-\pp+\QQ/2)\} = \epsilon(Q)/2 - 2\mu$; 
see Fig.~\ref{kinematics}. From the difference of $\Theta$-functions,
$k_F \le q \le q_{\rm max}(\omega) \equiv k_F\left[ 1 + |\omega|/\epsilon_F\right]^{1/2}$. 
This implies that $-|\omega| \le \Omega \le 0$. 
Together with the kinematical constraint $|k-q| \le Q \le k+q$, this leads to
$Q_{\rm min} = |k - q_{\rm max}(\omega)|$. For non-zero ${\rm Im}\Sigma$ we thus need the
kinematically allowed region (shaded rectangle in Fig.~\ref{kinematics}) to overlap
with $\Omega \ge \Omega_0(Q)$. This leads to the simple condition 
$Q_{\rm min} \le Q_0$, where the definition $\Omega(Q_0)=0$ leads to $Q_0 = 2k_F$. 
(We have also found, but do not discuss here, the $\omega >0$ threshold for $A \ne 0$.)

\begin{figure}
\centerline{\epsfxsize=7.0truecm \epsfbox{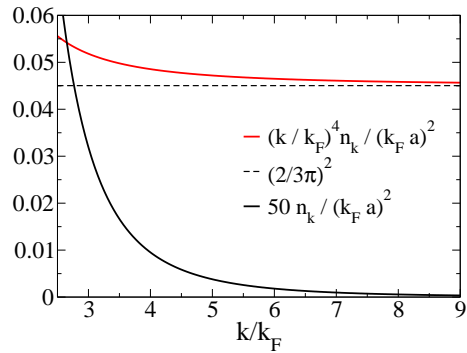}}
\caption{(color online) Momentum distribution tail for the dilute repulsive Fermi gas
with $na^3 = 3.4 \times 10^{-5}$.}
\label{nofk-tail}
\end{figure}

Solving $|k - q_{\rm max}(\omega)| = 2k_F$, we find 
$A(\kk,\omega <0) \ne 0$ in the range of energies $\omega = - \epsilon(\kk) - 3\epsilon_F \pm 2v_F k$; 
see Fig.~\ref{akw-galintskii}. For $k\gg k_F$, this simplifies to 
$|\omega+\epsilon(\kk)| \le 2v_F k$.  
Although the width of this range grows linearly with $k$, it becomes small 
relative to the central energy which grows like $-k^2$ for large $k$. 
We plot in Fig.~\ref{nofk-tail} the $n(\kk)$-tail
using $\int_{-\infty}^0 d\omega A(\kk,\omega)$ and find that
it agrees with the analytical result \cite{belyakov}. 
The incoherent spectral weight in $A(k \gg k_F,\omega)$ in the 
interval $|\omega+\epsilon(\kk)| \le 2v_F k$ is thus precisely $(2 k_F a / 3 \pi)^2 (k_F / k)^4$. 

%====================================================================================

{\bf Highly Imbalanced Fermi gas:}
We next turn to a two-component \emph{attractive} Fermi gas with
scattering length $a$ tuned through a broad Feshbach 
resonance \cite{zero-range}. 
While the ground state for equal spin populations is a superfluid exhibiting the
BCS-BEC crossover, we consider the different regime of large spin imbalance $n_\up/n_\dn$. 
There is by now considerable theoretical \cite{giorgini,villette}
and experimental evidence \cite{ketterle2008,zwierlein2009} that, for sufficiently large imbalance, 
superfluidity is destroyed for a large range of values of $a$, including unitarity $|a|=\infty$, and the
ground state is a (partially polarized) normal Landau Fermi liquid. 

For large $|a|$, we use the number of fermion species $2{\cal N}$ with
an $Sp(2{\cal N})$-invariant interaction as an artificial parameter to control the 
calculation in a large-${\cal N}$ expansion \cite{nikolic,villette}.
To first order in $1/{\cal N}$, ladder diagrams in the p-p channel determine
the self-energy. The resulting expressions are similar to
those used above and we show them schematically,
highlighting the differences that arise from spin-imbalance.
We now have $L = T \sum G^0_\up G^0_\dn$ where 
$G^0_\sigma(k) = 1/[ik_n - \xi_\sigma(\kk)]$ with
$\xi_\sigma(\kk) = \epsilon(\kk) - \mu_\sigma$.
The minority self-energy is given by $\Sigma_\dn = T \sum \Gamma G^0_\up$.
${\rm Im}\Sigma_\dn$ is then given by eq.~(\ref{imsigma})
with $\xi$ replaced by $\xi_\up$ both in the $\Theta$-function and in the definition of
$\Omega$.

We can analytically determine the energy range for which 
${\rm Im}\Sigma$, and hence $A$, are non-zero.
In Fig.~\ref{kinematics} we must now use $\Omega_0(Q) \equiv \epsilon(Q)/2 - 2\mu$
with $2\mu = \mu_\up + \mu_\dn$.
The final result \cite{stringent} is that, for $k\gg k_F$ and $\omega<0$,
$A(\kk,\omega)$ can be non-zero only in the range of energies
$|\omega+\epsilon(\kk)| \le \alpha v_{F\up} k$ where
$\alpha = \sqrt{2( 1 + \epsilon_{F\dn}/\epsilon_{F\up})}$. 

\begin{figure}
\centerline{\epsfxsize=8.5truecm \epsfbox{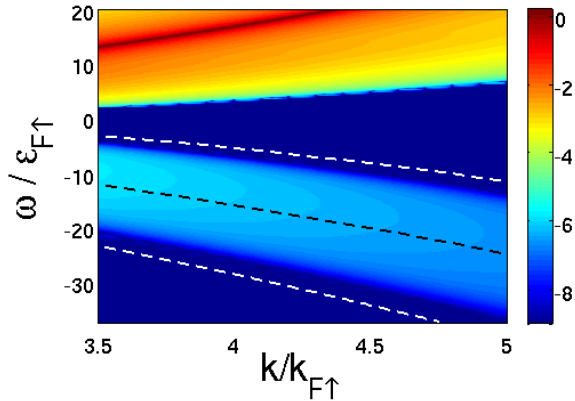}}
\caption{(color online) Logarithm intensity plot of $A_k (\omega)\epsilon_{F\up}$ 
for minority particles in the unitary Fermi gas
with imbalance $n_\dn/n_\up = 0.01$. The white dashed lines 
$\omega = - \epsilon(\kk) \pm \alpha v_{F\up} k$ are derived in the text; the black dashed line is
$\omega = - \epsilon(\kk)$. }
\label{akw-unitary}
\end{figure}

For concreteness, we focus here on unitarity $|a|= \infty$.
$A(\kk,\omega)$ for the highly imbalanced ($n_\dn/n_\up = 0.01$) unitary gas is shown in
Fig.~\ref{akw-unitary}. We have also verified that we get a $1/k^4$ tail for $n(\kk)$
in this system.
Our calculation of $A(\kk,\omega)$ is controlled only within
the $1/{\cal N}$-expansion. Note, however, that the singularity structure in the large $k$ limit
is determined only by short-distance properties of the two-body problem in vacuum,
while the strength of the singularity $C$ depends on the many-body state. 
The ladder approximation is exact for the
two-body problem. Thus we expect the bending back in the 
spectral function to be robust beyond the $1/{\cal N}$-expansion.

%====================================================================================
{\bf Superfluid State:} 
We now turn to a discussion of 
the superfluid ground state for a system with \emph{equal} densities of up and down spins and an
interaction described by a scattering length $a$. Unlike the normal Fermi liquids described above,
a branch of the dispersion that tracks $-\epsilon_\kk$ at large $k$ is very natural 
for the fermionic excitations in a superfluid \cite{haussmann}.
Nevertheless, even in this case, our analysis gives important quantitative insights.

In BCS mean field theory the spectral function 
$A_{\rm MF}(\kk,\omega) = v_\kk^2\delta(\omega + E(\kk)) + u_\kk^2\delta(\omega - E(\kk))$ 
where $v_\kk^2 = 1- u_\kk^2 = [1 - \xi(\kk)/E(\kk)]/2$. The excitation energy $E(\kk) = \sqrt{\xi^2(\kk) + \Delta^2}$
with $\Delta$ the energy gap.
For $k\gg k_F$, $E(\kk) \approx \epsilon(\kk)$ and $v_\kk^2 \approx \Delta^2/2\epsilon^2(\kk)$, so 
that $A_{\rm MF}(k\gg k_F,\omega<0) \approx [\Delta^2/2\epsilon^2(\kk)]\delta(\omega + \epsilon(\kk))$.
Thus we see that particle-hole mixing in the superfluid ground state naturally leads to a 
bending back of the dispersion.

However, there is a (large) quantitative problem with this result even in the BCS limit ($1/k_F a \ll -1$),
where one might have expected it to be the most accurate. Using $n(\kk) = \int_{-\infty}^0 d\omega A(\kk,\omega)$, 
or directly from BCS theory, one finds that the momentum distribution 
$n_{\rm MF}(\kk) = v_\kk^2 \approx \Delta^2/2\epsilon^2(\kk) = C_{\rm MF}/k^4$ 
for $k \gg k_F$. The problem is that the contact estimated from BCS theory 
$C_{\rm MF} \sim \Delta^2 \sim \exp(-1/k_F |a|)$ is exponentially small
in $|a|$. However, the exact answer \cite{tan,others} in the BCS limit is 
$C = 4\pi^2n^2a^2$ as $a \rightarrow 0^{-}$. To understand why BCS theory gets the wrong answer for $C$
we use the adiabatic relation \cite{tan}
$d{\cal E}/da = \hbar^2 C/(4\pi m a^2)$. As shown in \cite{dsr}, interaction effects
lead to  power-law corrections in $|a|$ in the ground state energy density ${\cal E}$,
which are numerically much more important than the essentially singular corrections coming from pairing.
In the extreme BCS limit, the contact is dominated by the Hartree term in ${\cal E}$ with calculable
corrections \cite{dsr}.

Thus the actual $A(k\gg k_F,\omega<0)$, 
even in the BCS limit, is dominated by interaction effects beyond BCS mean field theory. 
This results in a spectral weight
$C \sim |a|^2$ arising from interaction effects which exist even in the normal state, 
rather than resulting from pairing, which only makes an exponentially small contribution.  

%====================================================================================

{\bf Implications for RF spectroscopy:}
The physical effects we have discussed above lead to directly observable consequences in
RF spectroscopy experiments where an RF pulse is used to transfer atoms from one hyperfine level to another. 
The interpretation of these experiments is often complicated by two difficulties: the inhomogeneity of 
trapped gases and severe final state interactions. The first problem has been solved in the 
usual (``angle-integrated'') RF experiments using tomographic techniques. Final state
effects are not an issue in $^{40}$K \cite{jin2008}, and have been eliminated in $^6$Li by suitable choice
of hyperfine levels \cite{ketterle2008,zwierlein2009}. We emphasize that but for this
it would be very difficult to disentangle strong interactions in the many-body state
(self energy effects) from final state effects (vertex corrections) \cite{strinati}.
We will thus work in the (now experimentally relevant) limit where we ignore all
final state interactions.

Linear response theory then leads to the RF absorption intensity
$I_\sigma(\kk, \omega) = A_\sigma(\kk,\xi_\sigma(\kk) - \omega)f(\xi_\sigma(\kk) - \omega)$
where $\omega$ is the RF~shift.
The Fermi function $f(\omega)$ ensures that only occupied states can be excited by the probe.  
We set the multiplicative factor of the RF matrix element to unity so that
$\int d\omega \sum_{\kk} I_\sigma(\kk, \omega) = N_\sigma$.

Angle-resolved RF experiments \cite{jin2008} directly probe (the occupied part of) the
spectral function $A(\kk,\omega)$ and can see its unusual dispersion for $k\gg k_F,\ \omega<0$. 
As already noted in the Introduction, one cannot identify this bending back with 
pairing pseudogap physics \cite{varenna} since this universal feature also occurs in normal Fermi liquids.

The consequences of our results for the angle-averaged RF experiments, 
which measure $I_\sigma(\omega) = \sum_\kk I_\sigma(\kk, \omega)$, are more subtle. 
We now show that the unusual dispersion at large $\kk$ has a quantitative effect on the 
prefactor of the universal high-$\omega$ tail \cite{schneider} in $I_\sigma$. 
We rewrite $I_\sigma(\omega) = \sum_{\kk} \int_{-\infty}^0 d\Omega A_\sigma(\kk,\Omega)
\delta(\Omega - \xi_\sigma(\kk) + \omega)$ at $T=0$. In the $\omega\rightarrow\infty$ limit,
large negative $\Omega$ values, centered about $\Omega \approx - \epsilon(\kk)$, dominate. 
We thus find
$I_\sigma(\omega \to \infty) \simeq \sum_{\kk} n_\sigma(\kk)\delta(\omega - 2\epsilon_\kk)$.
Using \cite{tan} $n_\sigma(\kk) \approx C/k^4$ for $k \gg k_F$ we thus find that \cite{validity}
$I_\sigma(\omega \to \infty) \approx (C/{4 \pi^2 \sqrt{m}}) \omega^{-3/2}$.
The characteristic power-law is independent of the phase (normal or superfluid) of the Fermi gas,
though the value of $C$ does depend on the many-body state. 

The $\omega^{-3/2}$ tail has been discussed by various authors \cite{schneider,haussmann,strinati-b}.
First, we have derived this result analytically under very general conditions \cite{validity}. Second,
our results show that one must be very careful in interpreting its origin.
We emphasize that this tail and the bending back of the dispersion,
to which it is intimately related, arise from short-distance ``contact'' physics,
and should not by itself be taken as evidence for pairing.

{\bf Conclusions:}
We have shown that there is an unusual feature in the large momentum structure of the
single-particle spectral function of \emph{all} dilute Fermi gases, normal or superfluid,
which is closely related to the universal short distance features discussed by Tan and others
\cite{tan,others}. This is an incoherent branch of the dispersion where $\omega$ goes like
\emph{negative} $\epsilon_\kk$ \cite{combescot} that is quite unexpected in a \emph{normal} Fermi gas. 
Nevertheless, this is exactly what we find in the two systems where the ground state
is known to be a normal Landau Fermi liquid: the hard sphere Fermi gas and the highly imbalanced,
attractive Fermi gas. Even in a BCS superfluid, we show that this bending back at large $k$ is dominated by 
interaction effects rather than the pairing gap. 

{\bf Note Added:} Recently, we learned of a work by Combescot, Alzetto and Leyronas \cite{combescot},
where the approximation ${\rm Im} \Sigma(k \gg k_F, \omega <0) \propto \delta(\omega + \epsilon(\kk))$
is used which leads to a sharp feature in $A (\kk, \omega)$. While this may be sufficient for 
computing ``integrated'' quantities like $n(\kk)$, it
does not capture the incoherent structure in $A(\kk,\omega)$ described here.

We acknowledge discussions with D. Jin and E. Taylor 
and support from NSF-DMR 0706203 and ARO W911NF-08-1-0338.

%%%%%%%%%%%%%%%%%%%%%%%%%%%%%%%%%%%%%%%%%%%%%%%%%%%%%%%%%%%%%%%%%%%%%%%%%%%%%%%%%%%%%%%%%%%%%%%%%


\begin{thebibliography}{99}
\bibitem{range} 
For a hard sphere gas $a = r_0$, the range,
and the ``large'' $k$ regime where $n(\kk) \sim k^{-4}$ is $k_F \ll k \ll 1/r_0$.
Thus, unlike ref.~\cite{tan}, where $|a| \gg r_0$, we cannot set $r_0 = 0$.
\bibitem{mu-footnote}
We use $\mu = \epsilon_F$ in $L$ and $G^0$ since the corrections to $\mu$ are of order $k_F a$.
This is also done for the large-${\cal N}$ calculation for the polarized Fermi gas below.
\bibitem{KK} The simplification $\Gamma \approx g + g^2 L$ for the hard sphere gas
is valid only for $\omega \ll 1/ma^2$. ${\rm Im}\Sigma$ grows indefinitely for larger $\omega$.  
We subtract out this singular behavior
$S(z) = -i (2\sqrt{2}/3\pi) (k_F a)^2 \sqrt{\epsilon_F z}$
and then Kramers-Kronig transform 
$\mbox{Im} \widetilde{\Sigma} = \mbox{Im} \Sigma - \mbox{Im} S$.
The real self-energy is then $\mbox{Re} \Sigma  = \mbox{Re}
\widetilde{\Sigma} + \mbox{Re} S$.
\bibitem{checks}
We have checked our numerics against 
known results \cite{agd} for chemical potential $\mu$, quasiparticle residue $Z$,
effective mass $m^*$, and the scattering rate near the Fermi surface.
\bibitem{zero-range} Here $|a| \gg r_0$ and we
set the range $r_0 = 0$.
\bibitem{stringent}
We have obtained more stringent bounds for the imbalanced case,
using the detailed structure of ${\rm Im}\Gamma$, but the simpler 
analysis described here suffices
to establish a range linear in $k$ centered about $-\epsilon(\kk)$.
\bibitem{validity}
This large-$\omega$ tail is valid for $\epsilon_F \ll \omega \ll 1/ma^{\prime 2} \le 1/mr_0^2$
where $a^\prime$ characterizes the final state interactions and $r_0$ is the range.
\end{thebibliography}
\end{document}